# Processing slightly resolved ro-vibrational spectra during chemical vapor deposition of carbon materials: machine learning approach for plasma thermometry

R.R. Ismagilov,[1,a)] I.P. Kudarenko[1], S.A. Malykhin[1,2], S.D. Babin[1], A.B. Loginov[1], V.I. Kleshch[1], A.N. Obraztsov[1,2]

[1]Department of Physics, M.V. Lomonosov Moscow State University, Moscow 119991, Russia

[2]Department of Physics and Mathematics, University of Eastern Finland, Joensuu 80101, Finland

a) ismagil@polly.phys.msu.ru

**Abstract**

A fast optical spectroscopic method for determination rotational ($T_{rot}$) and vibrational ($T_{vib}$) temperatures in two-temperature Boltzmann distribution of the excited state by using machine learning approach is presented. The method is applied to estimate molecular gas temperatures in a direct current glow discharge in hydrogen-methane gas mixture during plasma-enhanced chemical vapor deposition of carbon film materials. Slightly resolved ro-vibrational optical emission spectrum of the $C_2$ ($v'=0 \rightarrow v''=0$) Swan band system was used for local temperature measurements in plasma ball. Random Forest algorithm of machine learning was explored for determination of temperature distribution maps. In addition to the $T_{rot}$, $T_{vib}$ maps, distribution maps and their gradients for electron temperature ($T_e$) and for the emission intensity of the spectral line 516,5 nm corresponding to $C_2$ species is presented and is discussed in detail.

**Introduction**

Non-thermal plasmas (also called as non-equilibrium plasmas, cold plasmas) are typically characterized by electron temperature $T_e$, vibrational temperature $T_{vib}$, rotational temperature $T_{rot}$ and translational temperature $T_{trans}$. The following inequalities commonly hold

$$T_e > T_{vib} > T_{rot} \simeq T_{trans} \qquad (1)$$

for the non-thermal plasmas generated by eternally applied constant electric field [1]. The non-equilibrium nature plays an important role in thin film deposition processes with participation of plasma activated species at moderate heating of the substrates. Various carbon film materials (diamond, graphene, carbon nanotubes etc.) may be produced, e.g., by direct current (DC) plasma-enhanced chemical vapor deposition (CVD) using hydrogen-methane gas mixture (see e.g. [2], [3] and refs there). Among the main parameters in the CVD processing the temperatures of substrate and activated gaseous media play crucial role [4]. The substrate temperature may be easily measured , e.g., by optical pyrometer, but determination of plasma temperature is much more challenging task. The rotational-vibrational (or, so called, ro-vibrational) optical emission spectrum of the plasma may be used as important source of information for temperature evaluation of its heavy particle components $T_{trans} \simeq T_{rot}$ [1]. The most common way for determination of the rotational temperature is based on "Boltzmann plotting" [5]. Since temperature dependence of the rotational line intensities follows to a Boltzmann distribution, the slope of the Boltzmann plot – logarithm of the reduced line intensity versus corresponding energy of the rotational transition – gives $1/kT_{rot}$, where $k$ is Boltzmann's constant, and $T_{rot}$ is the temperature should be determined [5]. However, there are several limitations in usage of this approach. The most prominent and commonly observed shortcoming is overlapping of the spectra leading to inaccuracies in the Boltzmann plots. In these cases, when Boltzmann plots are not applicable, rotational temperature determination is often done by using software packages that generate synthetic rotational spectra. These programs perform calculations of the spectra and use rotational and vibrational temperature as an input. Finding a temperature is done by comparison of the experimental spectrum with synthetic one. This procedure is often conducted by eye and has subjective character which is highly unwelcome in scientific practice.

The spectral dependencies necessary for temperature measurements are obtained usually via Optical Emission Spectroscopy (OES) and Laser Induced Fluorescence (LIF) [6]. The appropriate diatomic species should be chosen to be used as a thermometer depending on experimental circumstances [7]. In this work, we use OES as instrument for the temperature measurements and $C_2$ diatomic species as the most suitable source of information about environmental temperature. The ro-vibrational emission spectrum of the $C_2$ (v′=0 → v″=0) Swan band system is observed in differently activated carbonaceous gas media during formation of diamond, nanotubes, graphene and other carbon-related materials [8]–[11]. We present here, specifically, the results of the OES experimental study of the glow discharge during deposition of the carbon films containing the pyramidal diamond needles and carbon nanowalls (CNW). This study was conducted to determine a spatial distribution of the glow discharge parameters: a presence, relative concentration of $C_2$, their rotational and vibrational temperature profiles and an atomic hydrogen excitation temperatures.

**Experimental**

*2.1. Plasma-enhanced CVD systems and OES equipment*

Inspected plasma was ignited by a direct current (DC) glow discharge in hydrogen-methane gas mixture during plasma-enhanced chemical vapor deposition (PECVD). The detailed description of the experimental conditions and used CVD and OES setups are presented elsewhere [3]. Briefly, the DC discharge was activated in gas environment between a silicon substrate, located on a massive water cooled anode, and a similar water cooled cathode. The Si substrate temperature was measured by pyrometer (Cyclops, AMETEK Land). The deposition was performed using a gas mixture of methane and hydrogen with the constant total pressure 9.6 kPa. In this study two sets of the deposition parameters were used. The deposition parameters in one of the sets, noted further as '**case a**' were: $CH_4$ flow rate 14 sccm, $H_2$ flow rate 130 sccm, voltage drop 672 V and current 6.2 A, Si substrate temperature about 950 °C. These CVD parameters provide formation of the graphite-like materials [12]. Second set, noted further as '**case b**' was: $CH_4$ flow rate 5 sccm, $H_2$ flow rate 130 sccm, voltage drop 643 V and current 5.0 A, Si substrate temperature about 850 °C. These CVD parameters provide growth of the CVD diamond film containing the micrometer single diamond crystallites surrounded by the nanocrystalline and disordered carbon [13]. Small piece of Ge was situated nearby to Si substrates in the same way as it is commonly used for diamond doping purposes [14]. The anode (grounded) with Si substrate was located at bottom and the cathode above them.

Viewport with fused quartz window in the wall of the PECVD reactor chamber provides visual control of the activated gas region and OES detection. Imaging of the activated gas region on a screen (1:1 scale) located outside the reactor chamber was made by using a quartz objective. The screen was fixed on a motorized table which can be precisely moved in two directions (vertical and horizontal across optical axis). A throughout hole in the screen allows collection the portion of the light generated by the optically active species onto an entrance of the optical quartz fiber connected with a spectrometer (Hamamatsu Plasma Process Monitor C10346). The OES detection range spanned from 192 nm to 956 nm with separation on 1024 pixels. Typical optical emission spectrum measured with the system is shown in **Fig. 1A.** Default measurement parameters (exposure time - 25 ms, number of integration - 10) were applied and the emission lines were superficially analyzed with the build-in software (Hamamatsu Data acquisition software U9046). Deeper spectra analysis, control of movable screen and plasma-enhanced CVD monitoring were performed on a personal computer by using open source software [15]–[20].

*2.2. OES data processing*

*2.2.1. Spatial distribution map for particular emission lines and atomic hydrogen excitation temperatures*

Space distribution maps for chosen particular emission line were created using measured emission spectra. The temperature maps were obtained from analysis of the OES distributions.

In order to estimate electron temperature in sufficiently low pressure $CH_4/H_2$ plasma, the excitation temperature ($T_{exc}$) was calculates by using the relative intensities of the lines emitted by excited hydrogen atoms. We have assumed that the upper energy levels of the selected atomic transitions are in local thermodynamic equilibrium (LTE), that is, the population density of such levels follows the Boltzmann law. The latter allows us to use the conventional Boltzmann plot technique to determine $T_{exc}$ by using the expression

$$\ln\left(\frac{I_{ij}\lambda_{ij}}{g_i A_{ij}}\right) = \frac{-E_i}{kT_{ext}} + C \qquad (2)$$

where $I_{ij}$ is relative intensity of the emission line corresponding to transition between the energy levels *i* ang *j*, $\lambda_{ij}$ its wavelength, $g_i$ is the degeneracy or statistical weight of the emitting upper level *i* of the studied transition, and $A_{ij}$ is the transition probability for spontaneous radiative emission from the level *i* to the lower level *j*. Finally, $E_i$ is the excitation energy of level *i*, k is the Boltzmann constant and C is unimportant constant (see e.g. [21]). In this study the excitation temperature was estimated by using the intensity ratios of the Balmer lines. As it was mentioned previously [8] LTE does not maintained over the all inspected area in this PE CVD. In other words, in some parts upper levels of the Balmer α and β transitions are not thermalized with the free electrons in the plasma. The LTE assumption fails for such plasma, and the free electron temperature is not equivalent to the apparent excitation temperature obtained when a Boltzmann energy distribution is assumed for the atomic hydrogen excited states. For this reason the electron temperature $T_e$, measured using intensity ratios of the Balmer lines, may be considered only as a rough estimation. Nevertheless, these calculations make sense as an illustration of a general regularity in the free electron temperature distribution in plasma volume and its dependence on the conditions of the gas discharge. Hereafter, these terms $T_e$ and $T_{exc}$ will be used interchangeably.

*2.2.2. Machine learning approach for gas temperature estimations*

In this work, the commonly used experimental procedure mentioned in introduction section (related to eye comparison of experimental spectra with synthetic ones) was delegated to trained "artificial intelligence". This gas temperature estimation problem was considered as regression supervised learning problem, where program supposed to map input multidimentional variables from spectrometer to continuous $T_{rot}$ and $T_{vib}$ functions and predict (after training on synthetic spectra) latter values during real-time OES experiments. This prediction supposed to be sufficiently fast in order to provide temporal and spatial measurements with sufficiently good resolution, which impossible by using old-fasion eye-comparison methods.

Synthetic spectra ("Big Data") were generated under the assumption of two-temperature Boltzmann distribution description of the excited state, which might be calculated using classical approach described, e.g,. in [22]. The lines shape was tied to the instrumental spectroscopic function, which was calculated based on the FWHM of $H_\beta$ line (for Hamamatsu Plasma Process Monitor C10346 FWHM($H_\beta$) = 15 A). Noteworthy, another ready-to-use software might be found on the Internet (e.g. [23]) for $C_2$ Swan band synthetic spectra generation with reference to the instrument spectroscopic function. Examples of two synthetic OES spectra generated for $C_2$ (v′=0→v″=0) ro-vibronic band profile of Swan system in the case when $T_{rot}$ = 1310 K and $T_{vib}$ = 2450 K and in the case when $T_{rot}$ = 1310 K and $T_{vib}$ = 2450 K are presented in **Fig.1B** and **Fig.1C** respectively. 115521 spectra was generated, which cover rectangle grid with $T_{rot}$ expanding from 500 K to 6000 K (step 1 K) and $T_{vib}$ – 2000 K to 3000 K (step 50 K).

We divided the synthetic spectra randomly into a training set (80% of the all synthetic data) and a validation set (20% of the all synthetic data). In an effort to capture complicated relationships between the inputs and the outputs, we explore the effectiveness of the Random Forest (RF) [24] method, especially it's ability to handle high data dimensionality, being both fast and insensitive to overfitting. In this work RF classifier implemented in python module in scikit-learn package (version 21.3) [17] was used. In order to optimize RF performance the hyperparameters were tuned by using consequentially RandomizedSearchCV and GridSearchCV from scikit-learn tools. Optimized RF classifier were made of 1000 trees in the ensemble. The threshold for the minimum number of samples required to be at a leaf node was set to 1, and the threshold for the minimum number of samples required to split an internal node was set to 5. The maximum depth of the tree was set to 40. Other parameters were left as the defaults. With these values, stable test accuracy was observed with achieved $R^2$ (coefficient of determination) 0.9999989 for $T_{rot}$, and 0.999967 for $T_{vib}$. It should be noted, that at the time of writing this article Random Forest approach showed better results than XGBoost with $R^2$ [$T_{rot}$, $T_{vib}$] = (0.9999995, 0.9997) and Dense Neural Networks with achieved $R^2$ [$T_{rot}$, $T_{vib}$] = (0.9998, 0.9994).

We let the features be spectrometer channel intensity values. Forty spectrometer channels close to the $C_2$ (v′=0→v″=0) ro-vibronic band profile of Swan system ($d^3\Pi_g$-$a^3\Pi_u$) from 505.2 nm to 518.4 nm was excelled from each OES spectrum in order to use them as 40 features. As during training, every point in 40 dimensional features space is matched to the particular two-dimensional predicted *target* pairs ($T_{rot}$, $T_{vib}$).

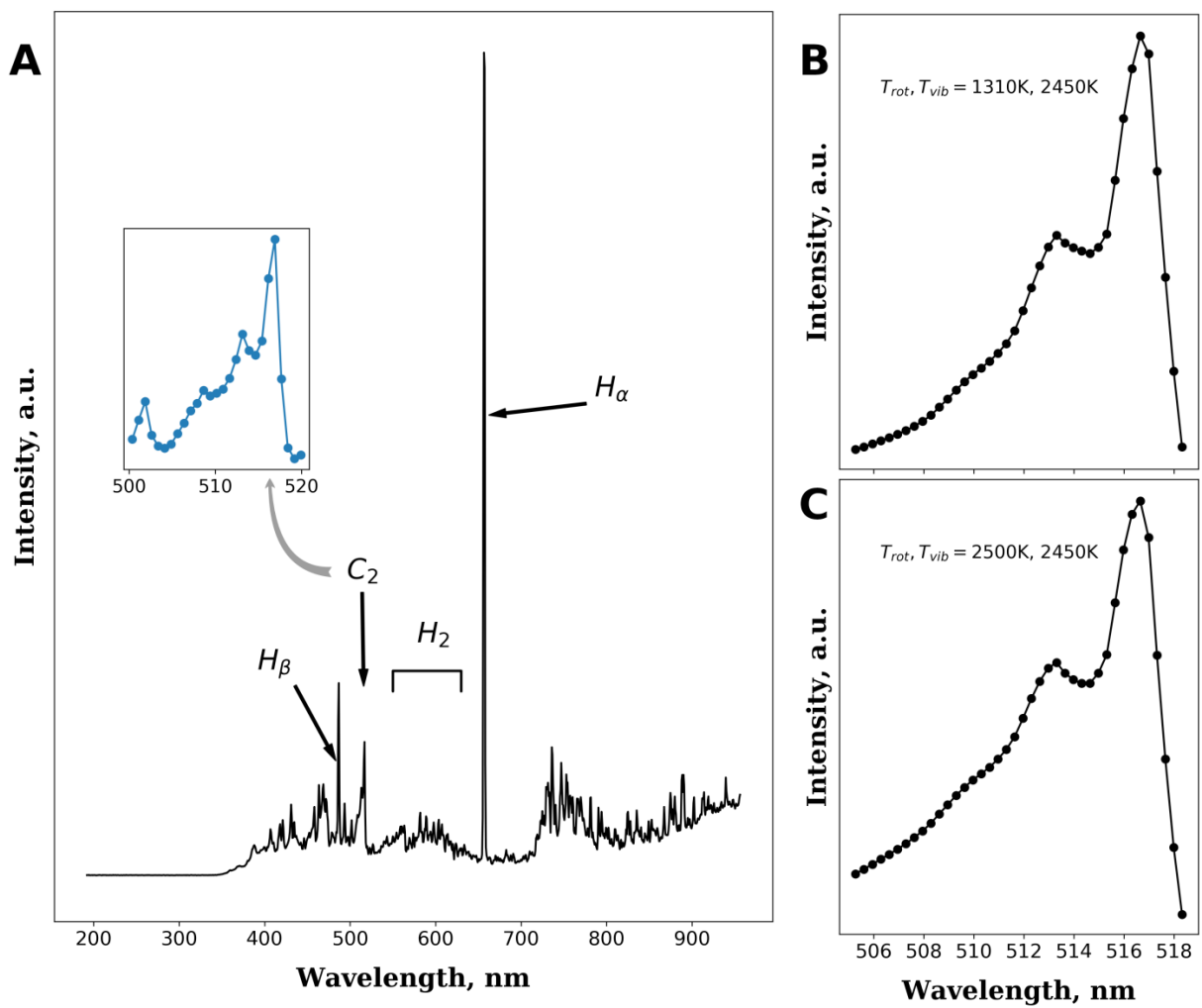

*Figure 1. (A) Typical OES spectrum collected by Hamamatsu Plasma Process Monitor C10346 from activated $H_2/CH_4$ gas mixture in plasma-enhanced CVD system. Indicated lines corresponds to species which appears and might be easily distinguished in any time and part of glowing discharge. (B, C) Examples of two synthetic OES spectra generated for $C_2$ (v'=0→v''=0) ro-vibronic band profile of Swan system in the case when $T_{rot}$ = 1310 K and $T_{vib}$ = 2450 K (B) and in the case when $T_{rot}$ = 1310 K and $T_{vib}$ = 2450 K (C). The total number of generated synthetic spectra is equal to 115521 and spectra covers rectangle grid with $T_{rot}$ expanding from 500 K to 6000 K (step 1 K) and $T_{vib}$ − 2000 K to 3000 K (step 50 K)*

**Results and discussions**

As an example **Fig. 1** shows a typical optical emission spectrum obtained for the plasma glowing in the hydrogen-methane gas mixture with 3.7% methane near 2.5 mm above the substrate surface in its central region. The Swan system with the Q(0, 0) band head at 516.5 nm as the most intense emission band attributed to $C_2$ and two atomic hydrogen emission lines at 656 nm ($H_\alpha$) and at 486 nm ($H_\beta$) of the Balmer transition series are clearly seen.

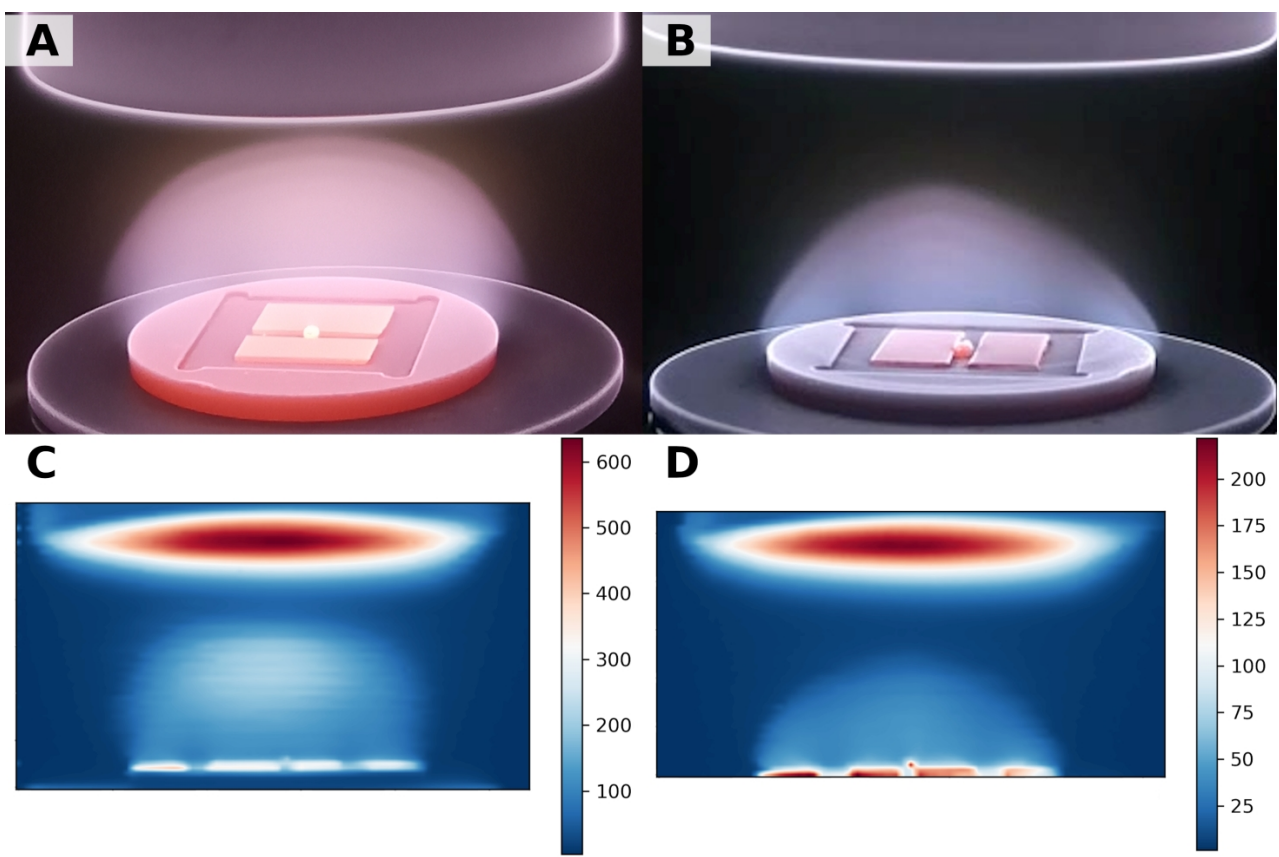

*Figure 2. (A, B) Typical photo images of $H_2/CH_4$ gas mixture activated zone for two distinct cases. The process parameters correspond to (A) carbon nanowalls (CNW) growth, (B) diamond needles growth. (C, D) A spatial distribution of the emission intensity in arbitrary units of the spectral line 516,5 nm corresponding to $C_2$ species, for two typical process parameters providing a deposition of carbon nanowalls (C) and diamond needles (D)*

In **Fig. 2A,B** typical photo images are presented for two considered CVD processes. It can be seen from photo images that the glow discharge manifests a stratification into dark and bright luminous layers. As expected the positive column (a luminous bell-like layer near the anode) is less bright than the negative glow (a thin luminous layer close to cathode). The positive column for both cases has a periodic layered structure composed of strations [25]. Close photo imaging of the cathode luminous layer also revealed it's layered nature. It is composed by two thin layer which supposed to be Cathode glow and Negative glow. The total thickness of these mentioned two glows are about 0.5 mm. Such thinness is explained by relatively high pressure conditions during CVD process, which appear to shrink these layers and pull them to the cathode (CVD p = 9.6 kPa ~ 70 Torr >> 0.1 Torr when a layered patterns extends to centimeters). It should be noted, that in **case b** luminous area at cathode covers electrode partly, while in **case a** there were no free surface and the glow covers the entire cathode surface, reaches every spot unprotected by dielectric. Hence, according to simplified classification [25] **case b** correspond to normal glow discharge, **case a** – to abnormal glow discharge. Detailed quantitative inspection of cathode region is saved for future works.

A spatial distributions of the emission intensity in arbitrary units of the spectral line 516,5 nm corresponding to $C_2$ species for two deposition cases are presented in **Fig. 2C,D.** These distribution maps nicely repeats the pattern and contours of corresponding photo images. They will be discussed in details further, but for now, it is worth to note, that since $C_2$ spectrum were chosen to be temperature source and trained Random Forest always gives pairs ($T_{rot}$, $T_{vib}$) even by feeding unreal data, these maps might be used as *trust* mask for revealing wrong temperatures, e.g. in **Fig. 2D** there are areas where $C_2$ intensity is too low and become close to the noise level, hence, predicted by Random Forest temperature supposed to be with low confidence.

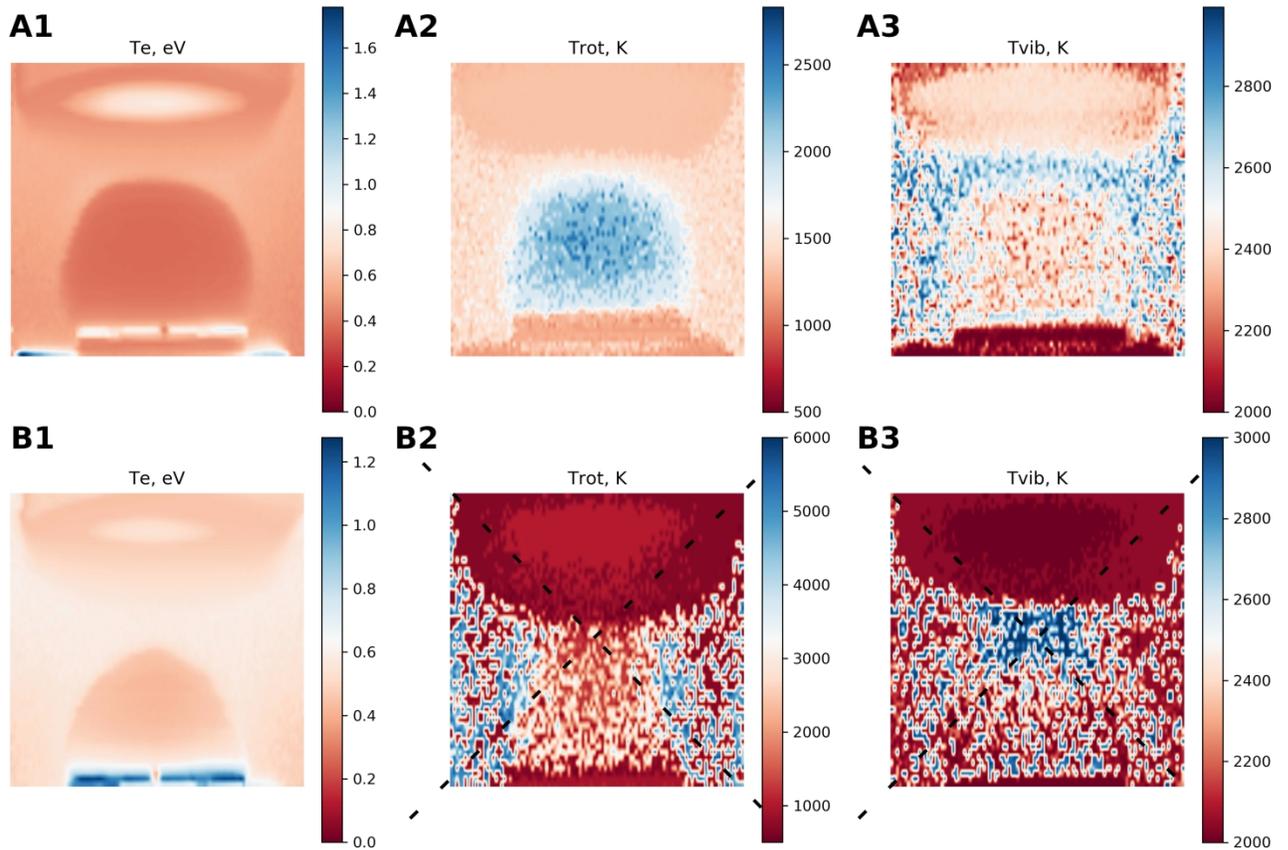

***Figure 3.*** *A spatial distributions of calculated $T_e$ by using intensity ratios of the Balmer lines and ($T_{rot}$, $T_{vib}$) by using Random Forest algorithm on experimental spectra of $C_2$ species, for two typical process parameters providing a deposition of carbon nanowalls (A) and diamond needles (B). 1 eV ~ 11 600 K*

A spatial distribution for $T_e$ calculated by using intensity ratios of the Balmer lines, and spatial distributions for $T_{rot}$ and $T_{vib}$ predicted by trained Random Forest after feeding it with experimental spectra for two deposition cases are presented in **Fig. 3.** At first glance, all distributions repeat the features which were seen in the photo images with it's the dark and bright luminous layers. However the $T_{rot}$ and $T_{vib}$ distribution maps for **case b** are distinguished by their low quality (see **Fig. 3B2,B3**) and huge temperature variation within Random Forest model training range. This is explained by the poor signal level of $C_2$ Swan band in some scanning parts, which were strongly expressed for **case b** with it's relatively low methane concentration. The fit temperatures are questionable due to the low signal level, but nonetheless, in some parts, e.g. along interelectrode axis, they show reasonable results. The temperature of solid substrates measured by pyrometer and $T_{rot}$ were very close (not only for presented here two deposition cases), indicating our right choice of using $C_2$ Swan band as the temperature source.

**Fig. 3** shows that electron temperature $T_e$ is about 0.5 - 1 eV(~5800 K - 11 600 K) and achieved the highest values at the anode, which high value >1 eV hides the temperature distribution details in the area between electrodes. Nevertheless, comparisons of these $T_e$ and $T_{rot}$ distribution maps allow us to recognize interesting features, tendencies: the electron temperature $T_e$ decreases progressively as one move from outside to inside the bell-shaped area, while the gas temperature $T_{rot}$ shows the opposite tendency and the maximum temperature occurs close to the middle point.

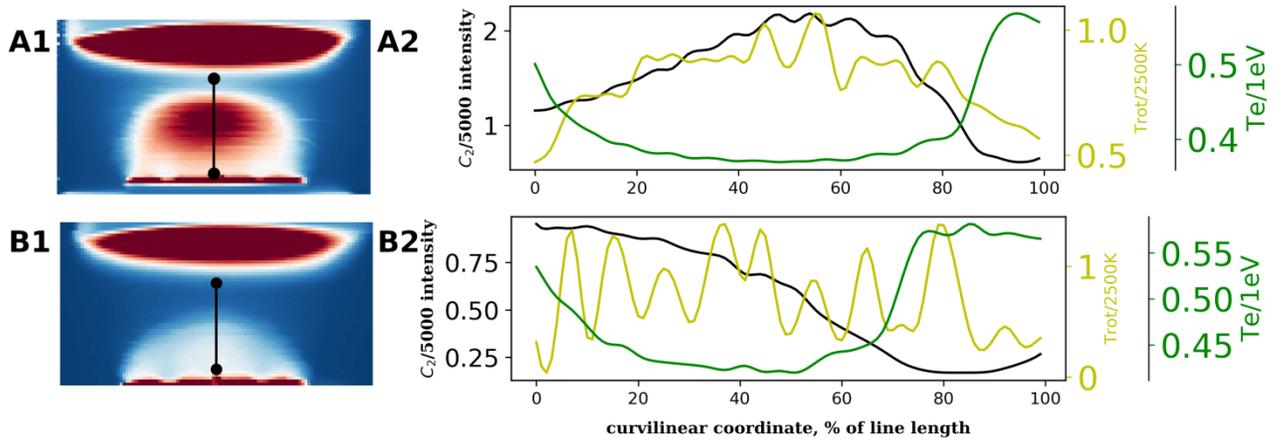

*Figure 4. A spatial distribution of the emission intensity of the spectral line 516,5 nm corresponding to $C_2$ species, for two typical CVD process parameters providing a deposition of carbon nanowalls (A1) and diamond needles (B1). Profiles in (A2, B2) indicate absolute intensity of the spectral line 516,5 nm [left axes], $T_{rot}$ scaled to 2500K [the first right axes] and $T_e$ in eV [the second right axes] along corresponding central, axial lines [from bottom to top, in % of total length] which depicted in (A1, B2). 1eV ~ 11 600K*

Several profiles including absolute intensity of the spectral line 516,5 nm (the most pronounced line corresponding to $C_2$ species), $T_{rot}$ and $T_e$ along central axial line is shown in **Fig.4.** In **case a** (see **Fig4. A1,A2**) the first two profiles have similar tendencies with pronounced global maximum. A similar dependence for $T_{rot}$ in the positive column was observed in atmospheric-pressure normal glow microplasmas in air [26] and were predicted by developed self-consistent two-dimensional computer fluid model of DC discharges in hydrogen and methane mixture under 50-200 Torr [27]. Latter third profile $T_e$, shows opposite to $T_{rot}$ variance with global not sharp and flat minimum, which position in comparison with previous extreme is a little closer to anode. In **case b s**imilar behavior observed only for $T_e$, while $T_{rot}$ oscillates around the mean ~1350 K and intensity of the spectral line 516,5 nm monotonously falls along the line.

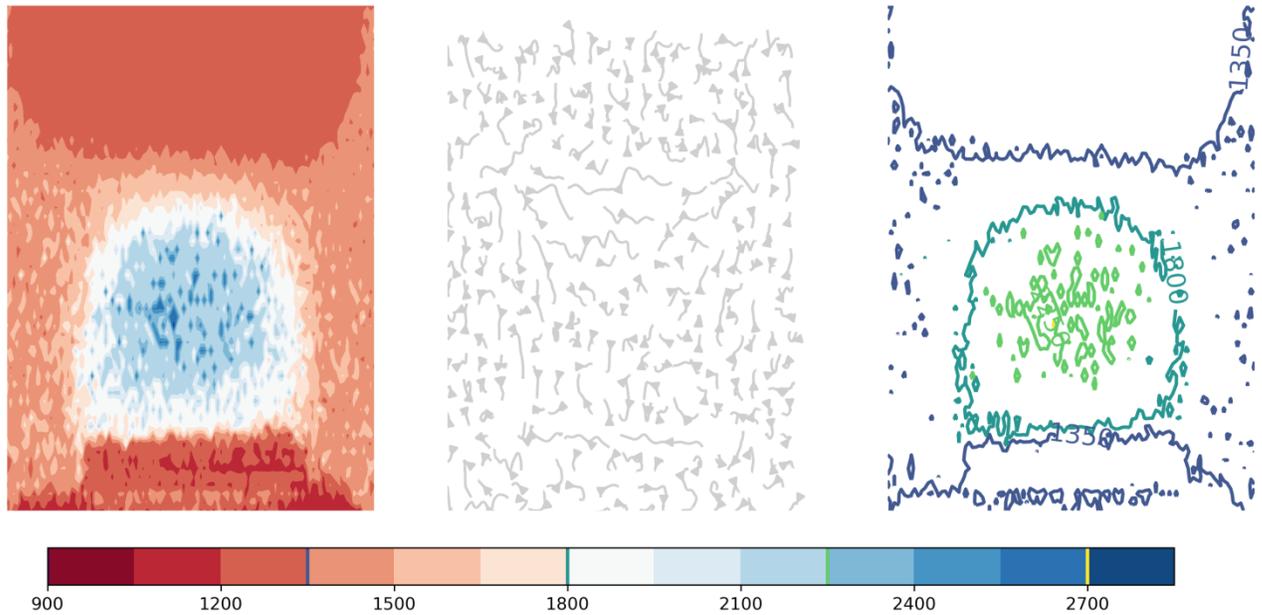

*Figure 5. (left) A $T_{rot}$ spatial distribution map predicted by Random Forest model in case when CVD parameters provide a deposition of carbon nanowalls, **case a**. (center) A "minus gradient" streamlines map for the left $T_{rot}$, $-grad(T_{rot})$. (right) contour plot for the same $T_{rot}$ map with constant levels: 1350 K, 1800 K, 2250 K, 2700 K.*

In addition to shown distribution maps of the discharge parameters, the gradients of corresponding maps related to flux are represented particular interest. Despite the fact, that total flux determined by summing up individual flux terms and expressed in rather complicated way, as a rule, through (minus) gradients of various quantities, e.g. including (minus) temperature gradients, (minus) concentration gradients (see for example Maxwell–Stefan diffusion equation [28]), further we consider streamline maps obtained by computing minus gradients of corresponding distribution maps.

**Fig. 5** shows spatial distributions with contours for $T_{rot}$ and corresponding white streamlines map (minus gradient $T_{rot}$). Streamlines map composed of randomly oriented short directional white lines with no concrete pattern, however it's natural to expect on average to get the outgoing flow from the hot center of the bell-shaped area in all directions to the colder areas. Such randomness probably caused by oscillations discussed earlier and showed in **Fig. 4.** It should be noted, that contour levels 1350 K (electrodes contours) and 1800 K (outer shell of bell-shaped area) might be drawn easily, while 2250 K, 2700 K can not be easily drawn because of their "random" and local appearance. If one increases contour level quantity and tries to add linear contours, many unrelated areas appears (that do not represent simply connected set) and image becomes clumsy. The situation become better for $C_2$, $T_e$ distributions (see **Fig. 6** and **Fig. 7**).

**Fig. 6** shows spatial distributions with contours for $T_e$ and corresponding white streamlines map (minus gradient $T_e$) for both considered cases. The anode region with it's high $T_e$ (see **Fig. 3**) was excluded from consideration in **Fig. 6** in order to reveal distribution details in the area between electrodes for electron temperature $T_e$. As expected from consideration $T_e$ profile along axial line (**Fig. 4**), the lowest $T_e$ value in the interelectrode gap achieved in a bell-shaped area. It was interesting to find out in **Fig. 6A1** (outlined by contour level 0.58 eV) that closed area, adjacent to the top-right and top-left to the bell-shaped area, exhibits the presence of local $T_e$ temperature increase. It is possible that such a closed area formation marks the beginning of the discharge transition to a state in which distinct "orange-yellow" layer forms under the bell-shaped area, composed presumably by heated dust particles, which have a continuous spectrum and were observed at the same places and described earlier [3]. Such layer usually found in cases with high methane concentration (more than 15%). It should be noted, that in considered cases methane concentration in gas mixture was higher for **case a**, but less than empirical critical level 15%. For the **case a** there was not any traces of continuous spectrum from a condensed substance heated up to a high temperature. Nevertheless, we were able to identify and mark this place.

Streamlines maps for both cases form specific but similar pattern, with only one noticeable difference for **case b** (see **Fig. 6B2**): there is a noisy layer composed of randomly oriented short directional white lines. This layer is matching with glow discharge dark area and it's real structure is questionable due to the relatively low OES signal level, but nonetheless, might represent reasonable results.

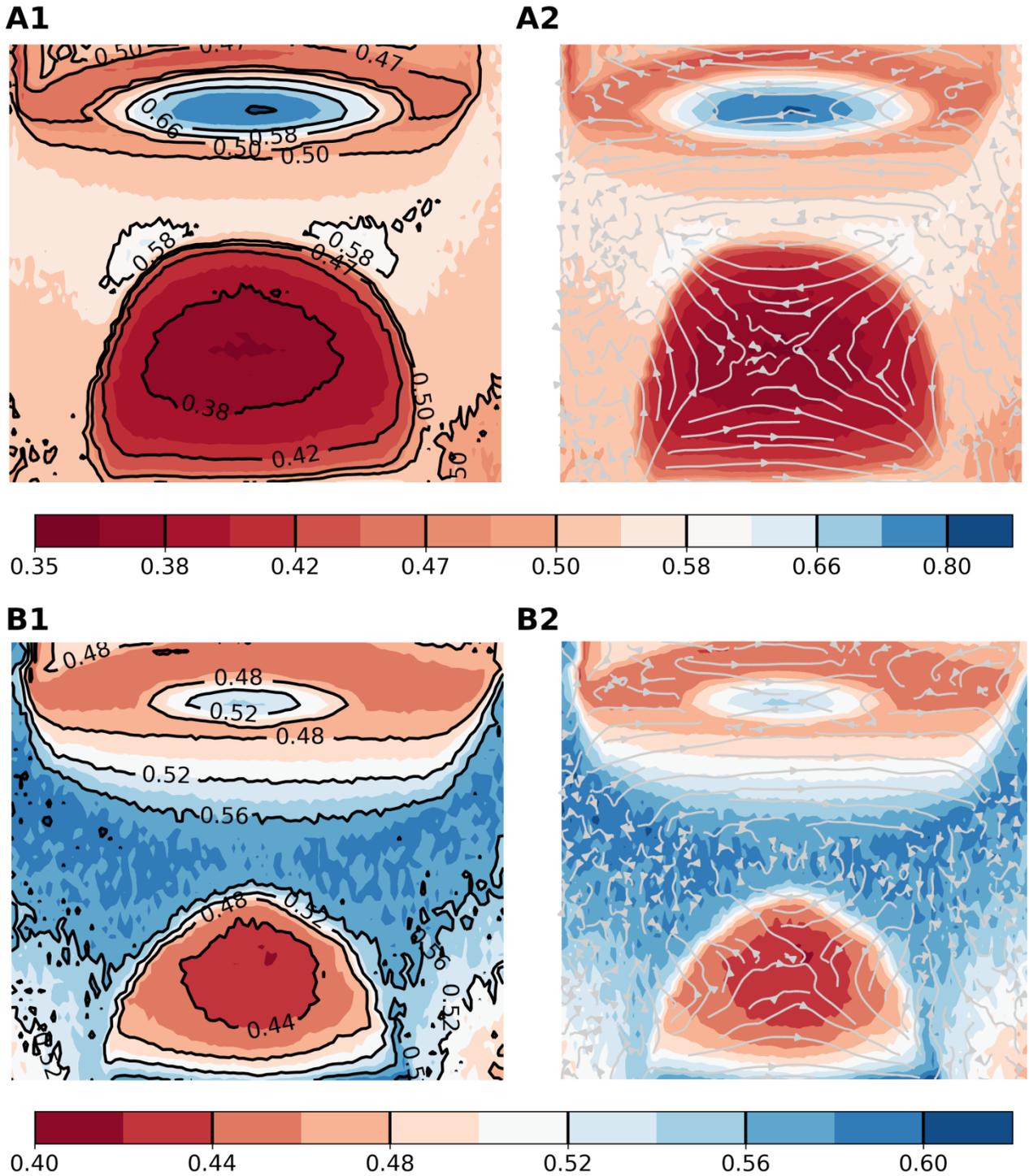

*Figure 6.* A spatial distribution of calculated $T_e$ with contours (A1, B1) and $T_e$ with "minus gradient" streamlines map for the corresponding $T_e$ distribution, $-grad(T_e)$ (A2, B2) for two typical CVD process parameters providing a deposition of carbon nanowalls (A1, A2) and diamond needles (B1, B2). Color bar indicates $T_e$ in eV, 1 eV ~ 11 600 K

Competing CVD system with similar in some extend characteristics to the considered one is plasma-enhanced CVD working at microwave range. Although optical emission intensity generally is not an accurate quantitative diagnostic for gas phase species concentrations, it has been shown [29], that the emission intensity of the (0,0) vibration band of the $C_2$ Swan system in $CH_4 + H_2$ plasma for microwave CVD correlates with the absolute $C_2$ concentration. With these weak connection in mind one might, for some extend, interpret distributions map shown in **Fig. 7** (without electrodes) as a relative distribution maps for $C_2$ density which, by the way, are not contradict with numerical simulation models [27]. Taking into account this remark, calculated minus gradient of distribution map and shown as white streamlines take on physical meaning, related to flux term responsible for diffusion (see Fick's or Maxwell–Stefan diffusion equations).

**Fig. 7** shows 2D spatial distributions with contours for the emission intensity of the spectral line at 516,5 nm corresponding to $C_2$ species, which complements the previously considered one-dimensional distribution along the electrode axis (see **Fig. 4**). In **case a** intensity of the line has pronounced global maximum in bell-shaped area, while in **case b** the distribution look like the same bell-shaped area sank lower with its maximum into the anode and only it's upper half protrudes. This phenomenon of immersion was also reflected in the distinct and smooth flux field pattern (compare white streamlines in **Fig. 7**).

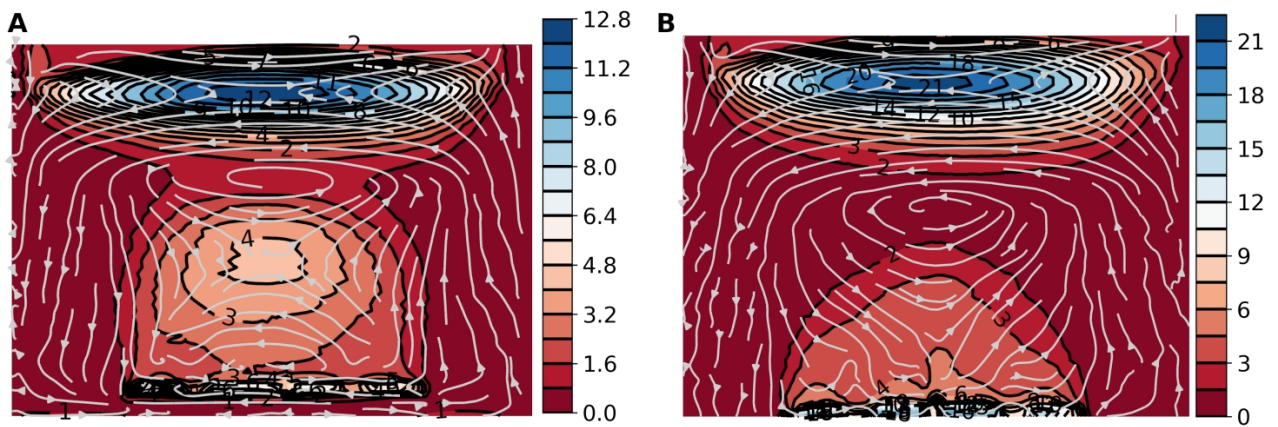

*Figure 7. A spatial distributions with contours for the emission intensity of the spectral line 516,5 nm corresponding to $C_2$ species and minus gradient (white "streamlines") of this distributions, for two typical CVD process parameters providing a deposition of carbon nanowalls (A) and diamond needles (B). Note the measured absolute intensities are divided by 5000 measured units (for A) and by 1000 measured units (for B). Hence, intensities for (A) several times higher than (B).*

## Conclusion

In this work Random Forest algorithm was used for quick and non-subjective comparisons of synthetic and experimental ro-vibrational emission spectra of $C_2$ diatomic species to estimate rotational $T_{rot}$ and vibrational $T_{vib}$ temperatures. Obtained spectra in different scanning area between electrodes in chemically reactive dc discharge are combined to form a composite temperatures map for activated $CH_4 + H_2$ gas mixture. Two CVD dc discharges considered in this study. First dc discharge led to the carbon nanowalls formation, second – the diamond films, containing the pyramidal diamond crystallites. In addition to the $T_{rot}$, $T_{vib}$ maps, distribution maps and their gradients for $T_e$ and for the emission intensity of the spectral line 516,5 nm corresponding to $C_2$ species is presented. Obtained results do not contradict the classical theory of cold plasma and complements with details the reached to date knowledge about it.

## Acknowledgments

The work was supported by RSF project 19-79-00203 and by RFBR grant #18-29-19071 (CVD, Optical parts and Si purchase). RRI and IPK are also grateful for support from the project MK-2046.2019.3.

## Data and code availability

All data, including python scripts, relevant to the current study are available from RRI, IPK on request. Initial scripts (obsolete) used for experimental data engineering can be found at github.com/Irebri/plasmaOEStoolbox